\documentstyle[epsfig,psfig]{texas}

\begin{document}

\title{The Mass Function of an X-Ray Flux-Limited Sample of Galaxy Clusters}
\author{Thomas H.\ Reiprich and Hans B\"ohringer}
\address{Max-Planck-Institut f\"ur extraterrestrische Physik\\
85740 Garching, Germany\\
{\rm Email: reiprich@mpe.mpg.de, hxb@mpe.mpg.de}}

\begin{abstract}
Using the brightest clusters in the ROSAT All-Sky Survey we compiled
an X-ray flux-limited sample of galaxy clusters. The clusters have
been reanalysed using ROSAT PSPC pointed observations if possible. The
gravitational mass has been determined individually for each cluster
in a homogeneous way assuming hydrostatic equilibrium. The mass
function has been derived. We present the preliminary results and a
comparison to previous determinations.
\end{abstract}

\section{Introduction}

Distribution functions of physical parameters of galaxy clusters can
place important constraints on cosmological scenarios. Comparison of the
mass function with
analytical or numerical calculations can yield, e.\,g., the amplitude
of the initial density fluctuations. Comparison of the X-ray luminosity,
gas temperature or mass function in different redshift bins can give
information about the cluster evolution.

Several authors have published an X-ray luminosity function, e.\,g.,
\cite{g90}, \cite{h92}, \cite{dg96}, \cite{e97}, \cite{r98}. Also a cluster
temperature function has been determined, e.\,g., \cite{h91},
\cite{m98}. A cluster gas mass function has been given by Burns et al.\
(1996), \cite{b96}, for
an optically selected cluster sample. A gravitational mass function has
previously been determined by Bahcall \& Cen (1993), \cite{bc93},
Biviano et al. (1993), \cite{b93}, and Girardi et al.\ \cite{g98}.
Bahcall \& Cen used the galaxy richness and
velocity dispersion to relate to cluster masses from the optical side and
a temperature-mass relation to convert the temperature function of Henry
\& Arnaud (1991), \cite{h91}, 
to a mass function from the X-ray side. Biviano et al.\ and Girardi et al.\
 used
velocity dispersions for an optically selected sample to determine the mass
function.

Since there is now high quality X-ray data available for the sample
selection --
using the ROSAT All-Sky Survey (RASS) -- and detailed cluster
analysis -- using ROSAT and ASCA pointed observations -- we have derived for
the first time the galaxy cluster gravitational mass function using
individually determined X-ray masses.

\emph{This article will be published
in the Proceedings of the 19$^{th}$ Texas Symposium on
Relativistic Astrophysics, held in Paris (1998), and is also available at:
http://www.xray.mpe.mpg.de/$\sim$reiprich/act/publi.html}

\section{The Sample}

Completeness of a cluster sample is essential for the
construction of the mass function. We compiled the clusters from RASS-based
cluster surveys of high completeness (REFLEX, NORAS \cite{b99}) and compared
also with other published catalogs \cite{e98}, \cite{e96}, \cite{beu98},
\cite{l89}, \cite{e90}. To avoid the high absorption and the crowded stellar
field in which clusters are hardly recognized in the galactic plane, 
only clusters with a galactic latitude $|b|\ge 20.0$ have been included. For
the same reasons the area around the Magellanic Clouds has been excluded. In
addition the Virgo cluster region has been excluded here. The sky coverage
is 26,720 deg$^2$.

\begin{figure}
\centering
\psfig{file=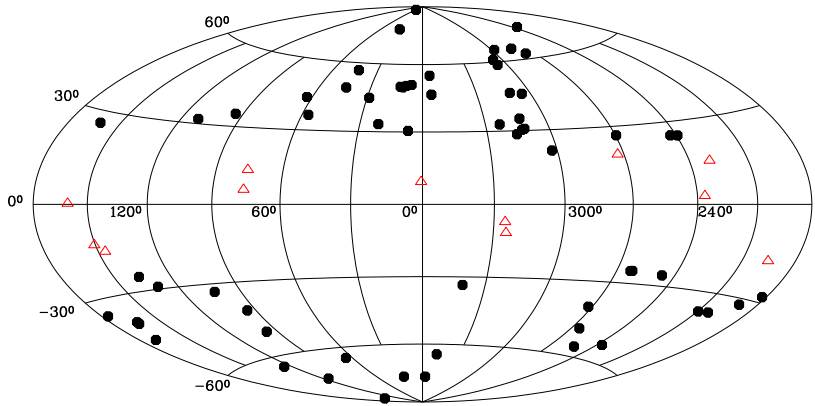,width=9cm,angle=270,clip=}
\caption{Distribution of HiFluGCS in galactic
coordinates. Open Triangles indicate clusters lying in the direction of
the galactic plane, these are not included in HiFluGCS.}\label{aito}
\end{figure}

We reanalysed the clusters using mainly ROSAT 
PSPC pointed observations and determined the X-ray flux
$f_{\rm X}${\small (0.1--2.4\,keV)}. 63 clusters have a flux greater
than or equal to our adopted flux limit
$f_{\rm X_{lim}}${\small (0.1--2.4\,keV)}$=2.0\cdot
10^{-11}\,\rm erg/s/cm^2$.
We call this cluster sample HiFluGCS (the Highest
X-ray Flux Galaxy Cluster Sample). The distribution of HiFluGCS in galactic
coordinates can be seen in Fig.\ \ref{aito}.\\ 
\section{Data Reduction and Analysis}
We used mainly high exposure ROSAT PSPC pointed observations to determine the
surface brightness profiles of the clusters, excluding obvious point sources.
If no pointed PSPC observations were
available in the archive or if clusters were too large for the field of view
of the PSPC we used RASS data. To calculate the gas density profile
the standard $\beta$-model \cite{c76}, \cite{g78} (equ.\ \ref{beta1}) has been
used.
\begin{equation}
\rho_{\rm gas}(r)=\rho_{\rm gas}(0)\left(1+\frac{r^{2}}{r_{\rm c}^{2}}\right)^{-\frac{3}{2}\beta}
\label{beta1}
\end{equation}
\begin{equation}
S_{\rm X}(R)=S_{\rm X}(0)\left(1+\frac{R^{2}}{r_{\rm c}^{2}}\right)^{-3\beta+\frac{1}{2}}
\label{beta2}
\end{equation}
Fitting the corresponding surface brightness formula (equ.\ \ref{beta2})
to the observed surface brightness profiles gives the parameters needed to derive
the gas density profile. To check if the often detected central
excess emission (central surface brightness of a cluster exceeding the fit
value) biases the mass determination we also fitted a double $\beta$-model of the
form $S_{\rm X}=S_{\rm X_1}+S_{\rm X_2}$ and calculated the gas mass profile by
$\rho_{\rm gas}=\sqrt{\rho_{\rm gas_1}^2+\rho_{\rm gas_2}^2}$. Comparison of
the single and double $\beta$-model gas masses shows good agreement.

We compiled the values for the gas temperature from the literature,
giving preference to
temperatures measured by the ASCA satellite \cite{ma98}, \cite{f97}, \cite{e91},
\cite{d93}. For clusters for which we did not find a published temperature we used
the X-ray luminosity-temperature relation given by Markevitch (1998),
\cite{m98}.

Assuming hydrostatic equilibrium the gravitational masses for the clusters can
be determined. Plugging equ.\ \ref{beta1} into the hydrostatic equation and
assuming the intracluster gas to be isothermal yields the gravitational mass
profile
\begin{equation}
M_{\rm tot}(r)=\frac{3kT_{\rm gas}r^{3}\beta}{\mu m_{\rm
p}G}\left(\frac{1}{r^2+r_{\rm c}^2}\right).
\label{hy}
\end{equation}

Having aquired the gravitational mass profiles for the clusters it is now
important to determine the radius at which to determine the cluster mass.
Simulations by Evrard et al.\ (1996), 
\cite{ev96}, have shown that the assumption of hydrostatic
equilibrium is generally valid within a radius where the mean gravitational
mass density is greater than or equal to 500 times the critical density
$\rho_{\rm c}=4.7\cdot10^{-30}\rm g\,cm^{-3}$, as long as clusters undergoing
strong merger events are excluded. This radius we call $r_{500}$. We calculated
the gravitational mass at $r_{500}$ and also $r_{200}$ which is usually referred
to as the virial radius. Using these definitions of the outer radius instead of
a fixed
length also allows the uniform treatment of clusters of different size.
Using $r_{500}$ also saves us from an extrapolation much beyond the significantly
measured cluster emission in general.

\section{Results}

\begin{figure}
\centering
\psfig{file=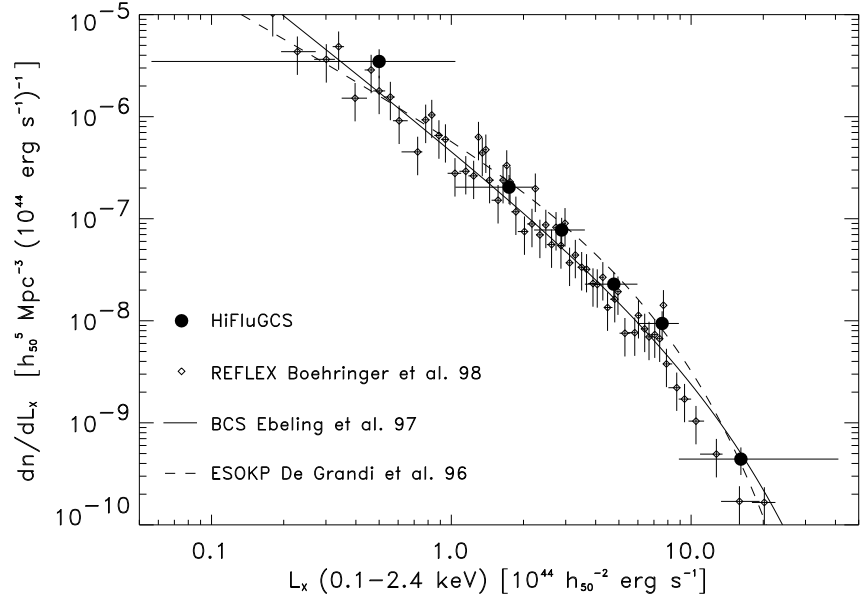,width=12cm,clip=}
\caption{Luminosity function of HiFluGCS compared
to luminosity functions of other cluster samples.}\label{lumi}
\end{figure}

In Fig.\ \ref{lumi} we show the X-ray luminosity function of HiFluGCS compared
to luminosity functions of other cluster samples. There is good agreement
between these determinations, if anything than HiFluGCS shows a marginally higher
density.
\begin{figure}
\centering
\psfig{file=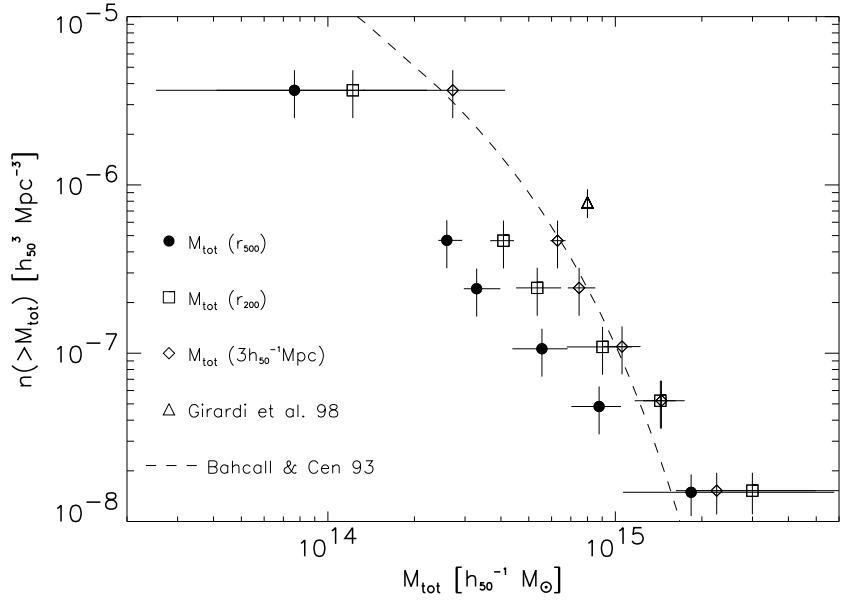,width=12cm,clip=}
\caption{Cumulative gravitational mass function of HiFluGCS for different
definitions of
the outer radius compared to previous determinations which used a fixed radius of 
$3\,h_{50}^{-1}\,\rm Mpc$.}\label{mass}
\end{figure}

In Fig.\ \ref{mass} we show the gravitational mass function for HiFluGCS 
for different definitions of the outer radius. Also shown are the mass functions
obtained by Bahcall \& Cen (1993), \cite{bc93}, and Girardi et al.\ (1998),
\cite{g98}, for an outer radius of $3\,h_{50}^{-1}\,\rm Mpc$.
Comparing our $r_{500}$ mass
function with that of Bahcall \& Cen we find increasing discrepancy towards
lower mass clusters up to a factor of 7--8. For the $r_{200}$ mass
function this discrepancy becomes less for the lower mass clusters but a
discrepancy arises towards the high mass end. In order to be able to directly
compare the mass function for the clusters in HiFluGCS with the two others shown
in Fig.\ \ref{mass} we determined the gravitational mass also at a fixed radius
of $3\,h_{50}^{-1}\,\rm Mpc$. Apart from the highest mass bin we almost exactly
reproduce the mass function determined by Bahcall \& Cen in this way.
The value given by Girardi et al.\ lying a factor 3--4 higher.

\section{Discussion}

Two major points are of concern when deriving the mass function:\\
1) The sample
completeness and 2) the reliability of the mass estimates. \\ \\
1) We compiled the clusters from RASS-based cluster surveys. These surveys
are complete at the 90\,\% level. The incompleteness of these surveys is
likely to be highest at fluxes close to their adopted flux limit, which is
much lower (factor $\sim 5$) than the flux limit adopted for HiFluGCS.
Additionally we checked further published X-ray cluster catalogs. Besides the
fact that still some clusters need to be checked in more detail we conclude
that HiFluGCS is essentially complete.\\ \\
2) For the first time we have determined a mass function with cluster masses
determined individually and in a homogeneous way for each cluster using
high quality X-ray data. However, galaxy clusters are generally not
spherically symmetric and hydrostatic equilibrium may not always be reached.
Simulations by Schindler (1996), \cite{s96}, and Evrard et al.\ (1996),
\cite{ev96}, however, have shown independently that the determined and true
mass do not differ dramatically ($\le 20\,\%$) if extreme merger clusters are
excluded. Clusters also may not always be isothermal. For instance
Markevitch et al.\
(1998), \cite{ma98}, find a general trend that the temperature decreases
with increasing radius, with the effect that the assumption of isothermality
leeds to an overestimation of the gravitational mass in the outer parts
($\sim 6$ core radii) of clusters by $\sim 30\,\%$. However, also average
relative temperature profiles of cluster samples have been found which
are consistent
with being isothermal, e.\,g., Irwin et al.\ (1999), \cite{i99}.

\section{Conclusions}

By reanalysing the brightest clusters of RASS-based galaxy cluster surveys
we constructed a complete X-ray flux-limited sample of galaxy clusters
(HiFluGCS),
the sky coverage being roughly 2/3 of the entire sky. We determined
global physical parameters for the clusters using mainly high exposure ROSAT
PSPC pointed observations. The luminosity function for HiFluGCS agrees
well with previous determinations. We determined the mass function 
for the first time by
individually determining the gravitational mass of each cluster in a
homogeneous way assuming
hydrostatic equilibrium and isothermality. Comparison
with previous determinations shows a strong discrepancy especially towards
lower mass clusters which is mainly due to the definition of the outer
radius. For comparison we also determined the cluster masses at a fixed radius
of $3\,h_{50}^{-1}\,\rm Mpc$ and apart from the highest mass bin we 
almost exactly reproduce the mass function determined by Bahcall \& Cen. 
However, we suggest to use as outer boundary a radius which depends on the
mean gravitational mass density, e.\,g.\, $r_{500}$, $r_{200}$, in order to
treat clusters of different size in a comparable way.

Another consequence of the definition of the outer radius becomes visible
when one integrates the mass function to determine the mass density bound in
galaxy clusters $\rho_{\rm bound}$. We find $\rho_{\rm bound}$ relative to
the critical density $\rho_{\rm c}$ to be 1.0\,\% for clusters of masses 
$2.5\cdot10^{13}\,h_{50}^{-1}\,\rm M_{\odot}$ and higher using $r_{500}$. The
fraction increases slightly to 1.6\,\% for clusters of masses 
$4.1\cdot10^{13}\,h_{50}^{-1}\,\rm M_{\odot}$ and higher when we use $r_{200}$.
We find a larger increase if we formally calculate the fraction for a fixed
radius of $3\,h_{50}^{-1}\,\rm Mpc$, which is 3.8\,\% for clusters of masses 
$1.3\cdot10^{14}\,h_{50}^{-1}\,\rm M_{\odot}$ and higher. Despite these different
results depending on the outer boundary, however, it is clear that only a
small portion of the total mass in the Universe is bound in galaxy clusters
as the largest collapsed entities, implying that most of the mass must
be somewhere else.

\section*{References}

\end{document}